\documentclass[a4paper,12pt]{article}
\usepackage[english]{babel}
\usepackage{graphicx,color}
\usepackage{amsmath,amsthm,amssymb,amsfonts}
\usepackage{amsthm}
\usepackage{mathtext}
\usepackage{exscale}
\usepackage{latexsym}
\usepackage[all]{xy}
\usepackage{epsfig}
\usepackage{lmodern}
\usepackage[T1]{fontenc}
\usepackage{textcomp}
\usepackage[cp1250]{inputenc}
\usepackage[bookmarks,colorlinks]{hyperref}

\hypersetup{colorlinks,%
linkcolor=red,%
urlcolor=green,%
citecolor=blue,%
filecolor=magenta}

\title{Macrostates thermodynamics and its stable classical limit in Global One--Dimensional Quantum General Relativity}

\author{
\textbf{L. A. Glinka\footnote{E-mail to: \href{mailto:glinka@theor.jinr.ru}{\texttt{glinka@theor.jinr.ru}}~,~\href{mailto:laglinka@gmail.com}{\texttt{laglinka@gmail.com}}}
}\vspace*{10pt}\\
\normalsize{\emph{Nicolai N. Bogoliubov Laboratory of Theoretical Physics}},\\
\normalsize{\emph{Joint Institute for Nuclear Research}},\\
\normalsize{\emph{141980 Dubna, Moscow Region, Russian Federation}}
}
\date{\today}

\newcommand{\tr}{\mathrm{Tr}}

\begin{document}

\maketitle
\thispagestyle{empty}
\begin{abstract}
Global One--Dimensional Quantum General Relativity is the toy model with nontrivial field theoretical content, describing classical one-dimensional massive bosonic fields related to any $3+1$ metric, where the dimension is a volume of three-dimensional embedding. In fact it constitutes the midisuperspatial Quantum Gravity model.

We use one-particle density operator method in order to building macrostates thermodynamics related with any $3+1$ metric. Taking the Boltzmann gas limit, which is given by the energy equipartition law for the Bose--Einstein gas of space quantum states generated from the Bogoliubov vacuum, we receive consistent with General Relativity thermodynamical degrees of freedom number.

It confirm that the proposed Quantum Gravity toy model has well-defined classical limit in accordance with classical gravity theory.

\end{abstract}
\newpage
\section{Introduction}
The toy model -- Global One--Dimensionality proposal in Quantum General Relativity -- considered in my last topical papers \cite{g1,g2,g3,g4,g5,g6}, is a nontrivial quantum field theoretical model describing one dimensional classical massive bosonic fields related immediately to $3+1$ decomposed metrics according to standard the Dirac--ADM approach in General Relativity. For construction of the model elementary quantum field theory methods, as field quantization by the Fock second quantization method and the Bogoliubov--Heisenberg diagonalization procedure, are used. In fact this simple type divagations constitutes a new and nontrivial midisuperspatial Quantum Gravity model, which results in space quantum states conception and unique connection between quantum correlations and physical scales of the system.

This paper is devoted to consider an application of one-particle density operator method in order to building thermodynamics of quantum macrostates related with any $3+1$ decomposed metric of General Relativity. Macrostates in the Quantum Gravity model are given by the Bose--Einstein gas of space quantum states. The self-consistence with General Relativity is achieved by the classical limit -- the Boltzmann gas limit of the macrostates thermodynamics, which is given by the energy equipartition law for classically stable phase of the Bose--Einstein gas of space quantum states generated from the Bogoliubov vacuum. The classically stable phase is defined by appropriate limit of quantum correlations for infinite number of vacuum space quantum states. In result we obtain classical degrees of freedom number which equal to number od space-time coordinates used in General Relativity.

The content of this paper is as follows. In the section 2 I recall the crucial elements of the Global One--Dimensional model of Quantum Gravity. There is present shortly a way from the Einstein--Hilbert General Relativity, by $3+1$ Arnowitt--Deser--Misner decomposition of metric and the Dirac primary quantization of the Hamiltonian constraint which lead to the Wheeler--DeWitt theory of quantum geometrodynamics, till my supposition about Global One--Dimensionality and field theoretical content of the Wheeler--DeWitt model. The section 3 is devoted to presentation of the main point of this article, that is macrostates thermodynamics and its classically stable limit. We use one-particle approximation. It is shown that the Boltzmann gas limit for classically stable configuration of the Bose--Einstein gas of macrostates generated from the stable Bogoliubov vacuum, leads to thermodynamical degrees of freedom number which is consistent with General Relativity.
\newpage
\section{Global 1D Quantum Gravity}
The classical gravity theory -- General Relativity -- describes 4-dimensional pseudo--Riemannian \cite{rie} differentiable manifold $(M,g)$ defined by metric $g_{\mu\nu}$ and coordinate system $x^{\mu}=(x^0,x^1,x^2,x^3)$, and characterized by the Christoffel connections $\Gamma^\rho_{\mu\nu}$, the Riemann curvature tensor $R^\lambda_{\mu\alpha\nu}$, the Ricci curvature tensor $R_{\mu\nu}$, and the scalar curvature $R$ \cite{krie,pet}
\begin{eqnarray}
&&\!\!\!\!\!\!\!\!\!\!\!\!\!ds^2=g_{\mu\nu}dx^{\mu}dx^{\nu},\qquad \Gamma^\rho_{\mu\nu}=\dfrac{1}{2}g^{\rho\sigma}\left(g_{\mu\sigma,\nu}+g_{\sigma\nu,\mu}-g_{\mu\nu,\sigma}\right)\\
&&\!\!\!\!\!\!\!\!\!\!\!\!\!R^\lambda_{\mu\alpha\nu}=\Gamma^\lambda_{\mu\nu,\alpha}-\Gamma^\lambda_{\mu\alpha,\nu}+\Gamma^\lambda_{\sigma\alpha}\Gamma^\sigma_{\mu\nu}-\Gamma^\lambda_{\sigma\nu}\Gamma^\sigma_{\mu\alpha},\quad R_{\mu\nu}=R^\lambda_{\mu\lambda\nu},\quad R=g^{\kappa\lambda}R_{\kappa\lambda}.
\end{eqnarray}
According to Einstein \cite{ein}, evolution of $(M,g)$ is given by the field equations\footnote{The units $8\pi G/3=c=\hbar=k_B=1$ are used.}
\begin{equation}\label{feq}
R_{\mu\nu}-\dfrac{1}{2}g_{\mu\nu}R+\Lambda g_{\mu\nu}=3T_{\mu\nu},
\end{equation}
where $\Lambda$ is cosmological constant, and $T_{\mu\nu}$ is Matter stress-energy tensor. The Einstein equations (\ref{feq}) can be received from the Hilbert dynamical action \cite{hil} modified by the Hartle--Hawking boundary $(\partial M,h)$ action \cite{hh}
\begin{equation}\label{eh0}
S[g]=\int_{M}d^4x\sqrt{-g}\left\{-\dfrac{1}{6}R+\dfrac{\Lambda}{3}+\mathcal{L}\right\}-\dfrac{1}{3}\int_{\partial M}d^3x\sqrt{h}K,
\end{equation}
where $K$ is extrinsic curvature of $(\partial M,h)$, by the Palatini principle \cite{pal} $\delta S[g]=0$ which relates the Matter Lagrangian $\mathcal{L}$ with $T_{\mu\nu}$
 \begin{equation}
   T_{\mu\nu}=\frac{2}{\sqrt{-g}}\frac{\delta\left(\sqrt{-g}\mathcal{L}\right)}{\delta g^{\mu\nu}}.
 \end{equation}

\subsection{3+1 Dirac--ADM approach}
By employing of the $3+1$ Dirac--ADM decomposition \cite{dir,adm,qft}
\begin{eqnarray}\label{dec}
g_{\mu\nu}=\left[\begin{array}{cc}-N^2+N^iN_i&N_j\\N_i&h_{ij}\end{array}\right], \qquad
g^{\mu\nu}=\left[\begin{array}{cc}-1/N^2&N^j/N^2\vspace*{5pt}\\
N^i/N^2&h^{ij}-N^iN^j/N^2\end{array}\right],
\end{eqnarray}
where $h_{ij}$, $N$, $N_i$ are embedding metric, lapse, shift functions, $h_{ik}h^{kj}=\delta_i^j$, $N^i=h^{ij}N_j$, the action (\ref{eh0}) takes the Hamiltonian form
\begin{eqnarray}\label{gd}
  S[g]=\int dt\int_{\partial M} d^3x\left\{\pi\dot{N}+\pi^i\dot{N_i}+\pi^{ij}\dot{h}_{ij}-NH-N_iH^i\right\},
\end{eqnarray}
where dot means $t$-differentiation, non vanishing conjugate momenta $\pi$'s are
\begin{equation}\label{mom}
\pi^{ij}=-\sqrt{h}\left(K^{ij}-h^{ij}K\right),
\end{equation}
and $H$, $H^i$ are defined as
\begin{eqnarray}
H=\sqrt{h}\left\{K^2-K_{ij}K^{ij}+ {^{(3)}R}-2\Lambda-6\varrho\right\}, \qquad H^i=-2\pi^{ij}_{~;j}~,\label{con2}
\end{eqnarray}
where ${^{(3)}R}=h^{ij}R_{ij}$ is scalar curvature of embedding and $\varrho=n^{\mu}n^{\nu}T_{\mu\nu}$ is energy density related to normal vector field $n^\mu=[1/N,-N^i/N]$ to a spacelike hypersurface. The Gauss--Codazzi equations \cite{gau,cod,han} determine the extrinsic curvature tensor $K_{ij}$ and extrinsic scalar curvature $K$ as
\begin{equation}
  K_{ij}=\dfrac{1}{2N}\left[N_{i|j}+N_{j|i}-\dot{h}_{ij}\right], \qquad K=\tr K_{ij},\label{con0}
\end{equation}
where stroke means intrinsic covariant differentiation. $H^i$ are diffeomorphisms $\widetilde{x}^i=x^i+\delta x^i$ generators
\begin{eqnarray}
i\left[h_{ij},\int_{\partial M}H_{a}\delta x^a d^3x\right]&=&-h_{ij,k}\delta x^k-h_{kj}\delta x^{k}_{~,i}-h_{ik}\delta x^{k}_{~,j}~~,\\
i\left[\pi_{ij},\int_{\partial M}H_{a}\delta x^a d^3x\right]&=&-\left(\pi_{ij}\delta x^k\right)_{,k}+\pi_{kj}\delta x^{i}_{~,k}+\pi_{ik}\delta x^{j}_{~,k}~~,
\end{eqnarray}
where $H_i=h_{ij}H^j$, and the DeWitt algebra \cite{dew}
\begin{eqnarray}
  i\left[\int_{\partial M}H\delta x_1d^3x,\int_{\partial M}H\delta x_2d^3x\right]\!\!&=&\!\!\int_{\partial M}H^a\left(\delta x_{1,a}\delta x_2-\delta x_1\delta x_{2,a}\right)d^3x,\label{com3}\\
  i\left[H_i(x),H_j(y)\right]\!\!&=&\!\!\int_{\partial M}H_{a}c^a_{ij}d^3z,\label{com1}\\
  i\left[H(x),H_i(y)\right]\!\!&=&\!\!H\delta^{(3)}_{,i}(x,y),\label{com2}
\end{eqnarray}
where $c^a_{ij}$ are structure constants of diffeomorphism group
\begin{equation}
  c^a_{ij}=\delta^a_i\delta^b_j\delta^{(3)}_{,b}(x,z)\delta^{(3)}(y,z)-\delta^a_j\delta^b_i\delta^{(3)}_{,b}(y,z)\delta^{(3)}(x,z)~~,
\end{equation}
is first-class type. Dirac's primary constraints time-preservation \cite{dew,dir1} leads to the secondary constraints (scalar and vector)
\begin{eqnarray}
\pi\approx0\rightarrow H\approx0, \qquad \pi^i\approx0\rightarrow H^i\approx0.
\end{eqnarray}
Vector constraint merely reflects spatial diffeoinvariance, scalar constraint gives dynamical information. Employing conjugate momenta (\ref{mom}) the scalar constraint becomes the Einstein--Hamilton--Jacobi equation \cite{ham1}--\cite{ham33}
\begin{equation}\label{con}
G_{ijkl}\pi^{ij}\pi^{kl}+\sqrt{h}\left(^{(3)}R-2\Lambda-6\varrho\right)=0,
\end{equation}
where $G_{ijkl}=\dfrac{1}{2}h^{-1/2}\left(h_{ik}h_{jl}+h_{il}h_{jk}-h_{ij}h_{kl}\right)$ is superspace metric.
\subsection{Quantum Geometrodynamics}
Canonical quantization \cite{dir,fad} of the Hamiltonian constraint (\ref{con})
\begin{eqnarray}\label{dpq}
i\left[\pi^{ij}(x),h_{kl}(y)\right]&=&\dfrac{1}{2}\left(\delta_{k}^{i}\delta_{l}^{j}+\delta_{l}^{i}\delta_{k}^{j}\right)\delta^{(3)}(x,y),\\
i\left[\pi^i(x),N_j(y)\right]&=&\delta^i_j\delta^{(3)}(x,y),\qquad i\left[\pi(x),N(y)\right]=\delta^{(3)}(x,y),
\end{eqnarray}
leads to the Wheeler--DeWitt equation \cite{whe, dew}
\begin{equation}\label{wdw}
\left\{-G_{ijkl}\dfrac{\delta^2}{\delta h_{ij}\delta h_{kl}}-h^{1/2}\left(-{^{(3)}R}+2\Lambda+6\varrho\right)\right\}\Psi[h_{ij},\phi]=0,
\end{equation}
where $\phi$ are Matter fields. Other first class constraints
\begin{equation}
  \pi\Psi[h_{ij},\phi]=0, \qquad \pi^i\Psi[h_{ij},\phi]=0, \qquad H^i\Psi[h_{ij},\phi]=0,
\end{equation}
merely reflects diffeoinvariance. The canonical commutation relations hold
\begin{equation}
\left[\pi(x),\pi^i(y)\right]=\left[\pi(x),H^i(y)\right]=\left[\pi^i(x),H^j(y)\right]=\left[\pi^i(x),H(y)\right]=0.
\end{equation}

\subsection{Global One--Dimensionality}
Supposing that Matter fields and the wave function $\Psi[h_{ij},\phi]$ are functionals of embedding's volume
\begin{equation}\label{GOD}
  \Psi[h_{ij},\phi]\rightarrow\Psi[h],~~h=\det h_{ij},
\end{equation}
and apply change of variables $h_{ij}\rightarrow\det h_{ij}$ in the Wheeler--DeWitt operator we obtain the Klein--Gordon--Fock type field equation for massive field $\Psi$
\begin{equation}\label{kgf}
\left(\dfrac{\delta^2}{\delta{h^2}}+m^2\right)\Psi=0,\qquad m^2=\dfrac{2}{3h}\left(^{(3)}R-2\Lambda-6\varrho\right),
\end{equation}
where $m^2$ is the mass square of $\Psi$. Elementary dimensional reduction of (\ref{kgf}) leads to the Clifford algebra and the Dirac type equation
\begin{equation}\label{dira}
  \left\{\mathbf{\Gamma}^{a},\mathbf{\Gamma}^{b}\right\}=2\eta^{ab}\mathbb{I},
  \qquad\eta^{ab}=\left[\begin{array}{cc}-1&0\\0&0\end{array}\right],\qquad\left(i\mathbf{\Gamma}\vec{\partial}-\mathbb{M}\right)\Phi=0.
\end{equation}
Here $\mathbf{\Gamma}=\left[-i\mathbb{I},\mathbb{O}\right]$ and we introduced notation
\begin{equation}
  \Phi=\left[\begin{array}{c}\Psi\\ \Pi_\Psi\end{array}\right], \qquad \vec{\partial}=\left[\begin{array}{c}\dfrac{\delta}{\delta h}\\0\end{array}\right], \qquad \mathbb{M}=\left[\begin{array}{cc}
0&1\\-m^{2}&0\end{array}\right]\geq0,
\end{equation}
where $\Pi_{\Psi}$ is conjugate momentum to $\Psi$ obtained from action $S[\Psi]$
\begin{equation}
  \Pi_{\Psi}=\dfrac{\delta S[\Psi]}{\delta \Psi}, \qquad S[\Psi]=-\dfrac{1}{2}\int\delta h\Psi^\dagger\left(\dfrac{\delta^2}{\delta{h^2}}+m^2\right)\Psi.
\end{equation}

\subsection{Field quantization}
Field quantization of (\ref{dira}) according to bosonic relations \cite{neu,a-w,bos}
\begin{eqnarray}
i\left[\mathbf{\Pi}_{\Psi}[h'],\mathbf{\Psi}[h]\right]=\delta(h'-h),\quad i\left[\mathbf{\Pi}_{\Psi}[h'],\mathbf{\Pi}_{\Psi}[h]\right]=0,\quad i\left[\mathbf{\Psi}[h'],\mathbf{\Psi}[h]\right]=0,\label{c3}
\end{eqnarray}
and the second quantization method \cite{ber,bs,bog} leads to the solution
\begin{equation}\label{sqx}
  \mathbf{\Phi}=\mathbb{Q}\mathfrak{B},\qquad\mathbb{Q}=\dfrac{1}{\sqrt{2}}\left[\begin{array}{cc}|m|^{-1/2}&|m|^{-1/2}\\
-i|m|^{1/2}&i|m|^{1/2}\end{array}\right].
\end{equation}
Here $\mathfrak{B}$ is a basis of creators $\mathsf{G}^{\dagger}[h]$ and annihilators $\mathsf{G}[h]$
\begin{equation}\label{db}
  \mathfrak{B}=\left\{\left[\begin{array}{c}\mathsf{G}[h]\\
\mathsf{G}^{\dagger}[h]\end{array}\right]:\left[\mathsf{G}[h'],\mathsf{G}^{\dagger}[h]\right]=\delta\left(h'-h\right),~~\left[\mathsf{G}[h'],\mathsf{G}[h]\right]=0\right\}.
\end{equation}
with dynamics determined by the system of equations
\begin{equation}\label{df}
\dfrac{\delta\mathfrak{B}}{\delta h}=\mathbb{L}\mathfrak{B},\qquad \mathbb{L}=\left[\begin{array}{cc}
-im&\dfrac{\delta}{\delta h}\ln\sqrt{|m|}\\
\dfrac{\delta}{\delta h}\ln\sqrt{|m|}&im\end{array}\right].
\end{equation}
Assuming new basis $\mathfrak{B}^\prime$ as compilation of the Bogoliubov transformation and the Heisenberg equations
\begin{eqnarray}
\mathfrak{B}^\prime=\left[\begin{array}{cc}u&v\\v^{\ast}&u^{\ast}\end{array}\right]\mathfrak{B}, \qquad |u|^2-|v|^2=1,\qquad \dfrac{\delta\mathfrak{B}^\prime}{\delta h}=\left[\begin{array}{cc}-i\omega&0\\0&i\omega\end{array}\right]\mathfrak{B}^\prime,\label{pr3}
\end{eqnarray}
where coefficients $u$, $v$ and frequency $\omega$ are functionals of $h$, gives the Bogoliubov coefficients dynamics
\begin{equation}\label{bcof}
  \dfrac{\delta\mathbf{b}}{\delta h}=\mathbb{L}\mathbf{b},\qquad \mathbf{b}=\left[\begin{array}{c}u\\v\end{array}\right],\qquad |u|^2-|v|^2=1,
\end{equation}
and the new static basis $\mathfrak{B}^\prime=\mathfrak{B}_{I}$ with stable vacuum $|0\rangle_I$
\begin{equation}\label{in}
\mathfrak{B}_{I}=\left\{\left[\begin{array}{c}\mathsf{G}_I\\
\mathsf{G}^{\dagger}_I\end{array}\right]: \left[\mathsf{G}_I,\mathsf{G}^{\dagger}_I\right]=1,~~\left[\mathsf{G}_I,\mathsf{G}_I\right]=0,~~\mathsf{G}_I|0\rangle_I=0\right\}.
 \end{equation}
Integration of (\ref{bcof}) can be done in the superfluid parametrization
\begin{eqnarray}\label{sup2}
u=e^{i\theta}\cosh\phi,\quad v=e^{-i\theta}\sinh\phi,\quad \theta=m_I\int_{h_I}^{h}\dfrac{\delta h'}{\lambda'},\quad\phi=-\ln{\sqrt{\left|\lambda\right|}},
\end{eqnarray}
where $\lambda=\dfrac{m_I}{m}=\dfrac{l}{l_I}$ scales sizes. By this reason we obtain finally
\begin{equation}\label{phi}
  \mathbf{\Phi}=\mathbb{Q}\mathbb{G}\mathfrak{B}_I,\qquad \mathbb{G}=\left[\begin{array}{cc}u^{\ast}&-v\\-v^{\ast}&u\end{array}\right],
\end{equation}
where $\mathbb{G}$ is the inverted Bogoliubov transformation matrix.

\subsection{Quantum correlations}
After quantization the equation (\ref{kgf}) can be rewritten in the form
\begin{equation}\label{qfe}
\left(\dfrac{\delta^2}{\delta h^2}+\dfrac{m^2_I}{\lambda^2}\right)\mathbf{\Psi}=0,
\end{equation}
and its solution can be red from (\ref{phi})
\begin{eqnarray}\label{field}
  \mathbf{\Psi}=\frac{\lambda}{2\sqrt{2m_I}}\left(\exp\left\{-im_I\int_{h_I}^h\dfrac{\delta h'}{\lambda'}\right\}\mathsf{G}_I+\exp\left\{im_I\int_{h_I}^h\dfrac{\delta h'}{\lambda'}\right\}\mathsf{G}_I^\dagger\right).
\end{eqnarray}
It is sensible to consider the many-field states acting on the vacuum
\begin{eqnarray}
|h,n\rangle\equiv\mathbf{\Psi}^n|0\rangle_I=\left(\frac{\lambda}{2\sqrt{2m_I}}e^{i\theta}\right)^n\mathsf{G}^{\dagger n}_I|0\rangle_I,
\end{eqnarray}
and determine the two-point quantum correlator $\langle n',h'|h,n\rangle$. In the normalization $\langle 1,h_I|h_I,1\rangle\equiv1$ the one-point correlator is fundamental
\begin{eqnarray}
    \langle 1,h|h,1\rangle=\lambda^2.\label{cor2}
\end{eqnarray}

\section{Macrostates thermodynamics}
\subsection{The Bose--Einstein gas}
The main point of this paper is macrostates thermodynamics and its classical stable limit. Field quantization with the stable Bogoliubov vacuum presented in the previous section allows to formulate formal thermodynamics of macrostates. We will use here one-particle approximation only.

In the one-particle approximation the density operator is number of states operator, which in static basis has a matrix $\mathbb{D}$ obtained in the Heisenberg--Von Neumann picture
\begin{equation}\label{do}
\mathsf{D}={\mathsf{G}}^{\dagger}{\mathsf{G}}=\mathfrak{B}^{\dagger}\left[\begin{array}{cc}1&0\\0&0\end{array}\right]\mathfrak{B}^{\dagger}=
\mathfrak{B}_I^{\dagger}\left[\begin{array}{cc}|u|^2&-uv\\-u^{\ast}v^{\ast}&|v|^2\end{array}\right]\mathfrak{B}_I
\equiv{\mathfrak{B}}_I^{\dagger}\mathbb{D}{\mathfrak{B}}_I.
\end{equation}
The number of states generated from the stable Bogoliubov vacuum is
\begin{equation}\label{nop}
 \xi=\dfrac{{_I}\langle0\left|{\mathsf{G}}^{\dagger}{\mathsf{G}}\right|0\rangle_I}{{_I}\langle0|0\rangle_I}=|v|^2,
\end{equation}
and using of elementary linear algebra methods allows to compute formal entropy  
\begin{equation}\label{entropia}
  S=-\dfrac{\tr\left(\mathbb{D}\ln\mathbb{D}\right)}{\tr\mathbb{D}}=\dfrac{8\xi(\xi+1)}{(2\xi+1)^2}-\ln\left(2\xi+1\right).
\end{equation}
Comparison of (\ref{entropia}) with the Bose--Einstein gas entropy \cite{ker} leads to the identification
\begin{eqnarray}
  2\xi+1\!\!&=&\!\!\exp\dfrac{U-\mu N}{T}-1,\label{id1}\\
  \dfrac{8\xi(\xi+1)}{(2\xi+1)^2}\!\!&=&\!\!\left(\dfrac{U-\mu N}{T}\right)\dfrac{\exp\dfrac{U-\mu N}{T}}{\exp\dfrac{U-\mu N}{T}-1},\label{id2}
\end{eqnarray}
which fix averaged number of states as
\begin{equation}\label{occu}
  N=\dfrac{1}{2\xi+1}.
\end{equation}
Taking the correct Hamiltonian matrix $\mathbb{H}$ of the Bose--Einstein gas
\begin{equation}
\mathsf{H}=\mathfrak{B}_I^{\dagger}\left[\!\!\begin{array}{cc}\dfrac{m}{2}\left(|v|^2+|u|^2\right)&-muv\\-mu^\ast v^\ast&\dfrac{m}{2}\left(|v|^2+|u|^2\right)\end{array}\!\!\right]\mathfrak{B}_I
\equiv{\mathfrak{B}}_I^{\dagger}\mathbb{H}{\mathfrak{B}}_I,
\end{equation}
One can compute internal energy and chemical potential according to standard rules
\begin{equation}\label{int1}
  U=\dfrac{\tr\mathbb{D}\mathbb{H}}{\tr\mathbb{D}},\qquad \mu=\dfrac{\delta U}{\delta N}.
\end{equation}

\subsection{Classically stable Boltzmann gas limit}
The second formula of (\ref{sup2}) and the number of states (\ref{nop}) allow to establish the relations for mass and size scales as well as for quantum correlations
\begin{eqnarray}\label{maas}
  \dfrac{m}{m_I}&=&\left(\sqrt{\xi}\pm\sqrt{\xi+1}\right)^2,\\
  \dfrac{l}{l_I}&=&\dfrac{1}{\left(\sqrt{\xi}\pm\sqrt{\xi+1}\right)^2},\\
  \langle1h|h1\rangle&=&\dfrac{1}{\left(\sqrt{\xi}\pm\sqrt{\xi+1}\right)^4}.
\end{eqnarray}
These formulas for the classical Boltzmann gas limit $\xi\rightarrow\infty$ becomes
\begin{eqnarray}
  \lim_{\xi\rightarrow\infty}\dfrac{m}{m_I}&=&\left\{\begin{array}{cc}\infty&~,~~\mathrm{for}~~$+$\\
  0&,~~\mathrm{for}~~$--$\end{array}\right.\\
  \lim_{\xi\rightarrow\infty}\dfrac{l}{l_I}&=&\left\{\begin{array}{cc}0&~,~~\mathrm{for}~~$+$\\
  \infty&,~~\mathrm{for}~~$--$\end{array}\right.\\
  \lim_{\xi\rightarrow\infty}\langle1h|h1\rangle&=&\left\{\begin{array}{cc}0&~,~~\mathrm{for}~~$+$\\
  \infty&,~~\mathrm{for}~~$--$\end{array}\right.
\end{eqnarray}
So it is clear the the classically stable physical object is obtained for the sign $-$. Computing for this case internal energy and temperature
\begin{eqnarray}\label{int2}
   U&=&m_I\dfrac{3\xi^2+3\xi+1}{2\xi+1}\left(\sqrt{\xi}-\sqrt{\xi+1}\right)^2,\\
  T&=&m_I\left[4\xi^2+4\xi+1-\dfrac{3\xi^2+3\xi+1}{\sqrt{\xi(\xi+1)}}(2\xi+1)\right]\dfrac{3\left(\sqrt{\xi}-\sqrt{\xi+1}\right)^2}{8\xi},
\end{eqnarray}
and using of the equipartition law according to the Boltzmann gas limit
\begin{eqnarray}\label{utr}
  \dfrac{U}{T}&=&\dfrac{\dfrac{8}{3}\dfrac{\xi}{2\xi+1}}{\dfrac{4\xi^2+4\xi+1}{3\xi^2+3\xi+1}-\dfrac{2\xi+1}{3\sqrt{\xi(\xi+1)}}},\\
\lim_{\xi\rightarrow\infty}\dfrac{U}{T}&=&\dfrac{f}{2},
\end{eqnarray}
leads to the number of thermodynamical degrees of freedom consistent with General Relativity
\begin{equation}
  f=4.
\end{equation}

\section{Conclusion}

This article was devoted to presentation of the next result of the Global One--Dimensionality model of Quantum General Relativity. There was recalled the idea of the model that is global change of variables $h_{ij}\rightarrow\det h_{ij}$ in the Wheeler--DeWitt equation and demanding that the Matter fields as well as effectively the Wheeler--DeWitt wave function are functionals of the global dimension. The model reduces 6 wave functions connected to 6 independent components of an embedding metric to 1 global wave function related to an embedding volume.

There was presented macrostates thermodynamics and its classically stable limit. The Bose--Einstein gas model was employed for computation of internal energy and temperature, and the Boltzmann gas limit was applied for the case of classically stable object, that is $l=\infty$ in the size scales. In result we have obtained the consistence with General Relativity - thermodynamical degrees of freedom number for the object is $f=4$, that lies in full agreement with the fact that space-time coordinates $x^{\mu}=(x^0,x^1,x^2,x^3)$ are considered as the degrees of freedom.

By this reason the presented model expresses nontrivial relation between the Einstein--Hilbert theory of the pseudo--Riemannian differentiable manifold and thermodynamics of macrostates generated from the stable Bogoliubov vacuuum, obtained by using the $3+1$ ADM decomposition of space-time metric and the Dirac--ADM canonical approach to General Relativity.

\section*{Acknowledgements}
The author benefited valuable discussions and excellent critical remarks from Profs. {A.B. Arbuzov}, {I.Ya. Aref'eva}, {B.M. Barbashov}, {K.A. Bronnikov}, {I.L. Buchdinder}, D.I. Kazakov, {V.N. Pervushin}, {V.B. Priezzhev}, and {D.V. Shirkov}.

\end{document}